\documentclass[superscriptaddress,reprint,prb,floatfix]{revtex4-1}
\usepackage{graphicx}
\usepackage{dcolumn}
\usepackage{amsmath,amssymb,amsfonts,amsthm,mathtools}
\usepackage{CJK,xparse,physics,color,bm}
\usepackage[colorlinks=true,breaklinks=true,allcolors=blue]{hyperref}

\providecommand{\ket}[1]{\ensuremath{\left|{#1}\right.\rangle}}
\providecommand{\bra}[1]{\ensuremath{\langle\left.{#1}\right|}}

\makeatletter
\DeclareFontFamily{OMX}{MnSymbolE}{}
\DeclareSymbolFont{MnLargeSymbols}{OMX}{MnSymbolE}{m}{n}
\SetSymbolFont{MnLargeSymbols}{bold}{OMX}{MnSymbolE}{b}{n}
\DeclareFontShape{OMX}{MnSymbolE}{m}{n}{
	<-6>	MnSymbolE5
	<6-7>	MnSymbolE6
	<7-8>	MnSymbolE7
	<8-9>	MnSymbolE8
	<9-10>	MnSymbolE9
	<10-12>	MnSymbolE10
	<12->	MnSymbolE12
}{}
\DeclareFontShape{OMX}{MnSymbolE}{b}{n}{
	<-6>	MnSymbolE-Bold5
	<6-7>	MnSymbolE-Bold6
	<7-8>	MnSymbolE-Bold7
	<8-9>	MnSymbolE-Bold8
	<9-10>	MnSymbolE-Bold9
	<10-12>	MnSymbolE-Bold10
	<12->	MnSymbolE-Bold12
}{}

\let\llangle\@undefined
\let\rrangle\@undefined
\DeclareMathDelimiter{\llangle}{\mathopen}{MnLargeSymbols}{'164}{MnLargeSymbols}{'164}
\DeclareMathDelimiter{\rrangle}{\mathclose}{MnLargeSymbols}{'171}{MnLargeSymbols}{'171}
\makeatother
\newcommand{\Bra}[1]{\left\llangle #1 \right|}
\newcommand{\Ket}[1]{\left| #1 \right\rrangle}
\newcommand{\Braket}[2]{\left\llangle #1 \vphantom{#2} \right|\kern-0.6ex\left. #2 \vphantom{#1} \right\rrangle}
\graphicspath{{.}{./Figures/}}

\begin{document}
\title{Hunting for the non-Hermitian exceptional points with fidelity susceptibility}
\begin{CJK*}{UTF8}{bsmi}
\author{Yu-Chin Tzeng}\email{yctzeng@mx.nthu.edu.tw}
\affiliation{Department of Physics, National Tsing Hua University, Hsinchu 30013, Taiwan}
\affiliation{Department of Applied Physics, Tunghai University, Taichung 40704, Taiwan}
\affiliation{Department of Physics, National Chung Hsing University, Taichung 40227, Taiwan}
\author{Chia-Yi Ju}
\email{cju@nchu.edu.tw}
\author{Guang-Yin Chen}
\author{Wen-Min Huang}
\affiliation{Department of Physics, National Chung Hsing University, Taichung 40227, Taiwan}

\date{\today}

\begin{abstract}
The fidelity susceptibility has been used to detect quantum phase transitions in the Hermitian quantum many-body systems over a decade, where the fidelity susceptibility density approaches $+\infty$ in the thermodynamic limits. Here 
the fidelity susceptibility $\chi$ is generalized to non-Hermitian quantum systems
by taking the geometric structure of the Hilbert space into consideration.
Instead of solving the metric equation of
motion from scratch, we chose a gauge where the fidelities are composed of biorthogonal eigenstates and can be worked out algebraically or numerically when not on the exceptional point (EP).
Due to the properties of the Hilbert space geometry at EP, we found that EP can be found when $\chi$ approaches $-\infty$.
As examples, we investigate the simplest $\mathcal{PT}$ symmetric $2\times2$ Hamiltonian with a single tuning parameter and 
the non-Hermitian Su-Schriffer-Heeger model. 
\end{abstract}
\maketitle
\end{CJK*}

\section{Introduction}
The overlap of two states, 
or fidelity $\mathcal{F}_\text{h}(\ket{\psi},\ket{\varphi})=\braket{\psi}{\varphi}\braket{\varphi}{\psi}$ to be more specific, is used
in the quantum information sciences as an estimation of the similarity of two quantum states. In quantum many-body system, the eigenstate fidelity and the fidelity susceptibility has been used to detect quantum phase transitions~\cite{fidelity_review}. Away from the critical point, two ground states with nearby parameters in the same phase are expected to have a high fidelity or a finite value of fidelity susceptibility. On the other hand, when the system is close to the quantum critical point, a quick drop in the fidelity or divergence in fidelity susceptibility density is very likely to occur. 
Although the definition of the fidelity is far from unique, e.g. the square root of $\mathcal{F}_\text{h}$ is also widely accepted, 
as a quantity that defines the closeness of two state, the fidelity satisfies the
Jozsa's axioms~\cite{Jozsa1994}: i) $0\leq\mathcal{F}_\text{h}(\ket{\psi},\ket{\varphi})\leq1$;
ii) $\mathcal{F}_\text{h}(\ket{\psi},\ket{\varphi})=\mathcal{F}_\text{h}(\ket{\varphi},\ket{\psi})$;
iii) $\mathcal{F}_\text{h}(U\ket{\psi},U\ket{\varphi})=\mathcal{F}_\text{h}(\ket{\psi},\ket{\varphi})$, where $U$ is an arbitrary unitary operator; and 
iv) $\mathcal{F}_\text{h}(\ket{\psi_1}\otimes\ket{\psi_2},\ket{\varphi_1}\otimes\ket{\varphi_2})=\mathcal{F}_\text{h}(\ket{\psi_1},\ket{\varphi_1})\mathcal{F}_\text{h}(\ket{\psi_2},\ket{\varphi_2})$.

Since the discovery of that a non-Hermitian Hamiltonian with parity-inversion ($\mathcal{P}$) plus time-reversal ($\mathcal{T}$) symmetry has real eigenvalues~\cite{Bender_1998}, it has been attracting both theoretical and experimental investigations on the non-Hermitian systems in many fields, ranging from classical optics, acoustics, optomechanics, mechanics, electronics, metamaterials, plasmonics, condensed matter physics, and photonic crystals to innovative devices~\cite{f1,f2,f3,f4,f5,f6,Motohiko_2019,f7,PTsym_xxz,f8, RNHQM, zhao2019non, PhysRevLett.122.185301, PhysRevResearch.2.033018, review2019}.
However, the fidelity $\mathcal{F}_\text{h}$ losses its meaning in the non-Hermitian quantum systems because the inner product in conventional quantum mechanics leads to all kinds of weird behaviors, e.g. faster-than-light communication~\cite{Lee2014}, and entanglement increasing under local operations and classical communications (LOCC)~\cite{Chen2014}. To generalize the fidelity for non-Hermitian systems, it is necessary to use a proper inner product. Utilizing the geometric meaning of Schr\"{o}dinger's equation, one find a proper inner product that is free from the above mentioned weird behavior. Rather than being a strictly Euclidean-like Hilbert space, the Hilbert space of the quantum states can have its own ``geometry'' or metric~\cite{metric,thermo,Ju2019}. 

Given a non-Hermitian Hamiltonian $H{\neq}H^\dagger$, the connection-compatible metric operator $G$, of the system has to be positive-definite, $G=G^\dagger$, and satisfy the equation of motion:
\begin{equation}\label{eq:eom}
        \frac{\partial}{\partial t}G=i(GH-H^\dagger G).
\end{equation}
The corresponding dual vector of $\ket{\psi}$ becomes $\Bra{\psi}{=}\bra{\psi}G$. For notation consistency, define $\Ket{\varphi}{=}\ket{\varphi}$ and the inner product becomes $\Braket{\psi}{\varphi}{=}\bra{\psi}G\ket{\varphi}$. The detailed derivation can be found in Ref.~[\onlinecite{Ju2019}]. It is known that if the Hamiltonian is not at the exceptional point, then the metric can always be chosen as
\begin{equation}\label{eq:metric}
        G=\sum_{i}\ket{L_i}\bra{L_i}.
\end{equation}
Where $\ket{L_i}$ is the $i$th left-eigenvector of the Hamiltonian, i.e. $\bra{L_i}H=\bra{L_i}E_i$. This gauge choice~\cite{Ju2019} not only reproduces the biorthogonal quantum mechanics~\cite{Brody_2013}, it also works the best with Hamiltonian eigenstates.
Note that the $j$th right-eigenvector $\ket{R_j}$ satisfies $H\ket{R_j}=E_j\ket{R_j}$ and the bi-orthonormal relation, $\braket{L_i}{R_j}=\delta_{ij}$.

The exceptional point (EP)~\cite{Kato1966,Heiss_2012,EP_science2019,EP_topology} of the non-Hermitian systems is a special point in the parameter space where both eigenvalues and eigenstates merge into only one value and state. In general, the eigenstates of a non-Hermitian Hamiltonian are not orthogonal with each other using the conventional inner product. As the parameter approaches the EP, two or more eigenstates close to each other and eventually coalesce into one. Thus, the Hamiltonian at the EP has less eigenvectors and can not span the entire Hilbert space. The EPs are also associated with the real-to-complex spectral transition for parity-time ($\mathcal{PT}$) symmetric Hamiltonians~\cite{Heiss_2012}.
The EPs provide genuine singularities, which can manifest prominently in optical~\cite{Nature2017}, plasmonic~\cite{plasmonic_EP} and microwave~\cite{doppler2016dynamically} response properties and hybrid dynamical systems~\cite{xu2016topological,zhang2017observation,PRL.121.137203}, cold atoms~\cite{Xu:17}, correlated systems~\cite{PhysRevB.99.121101,PhysRevB.100.115124}, sensing enhancement and scattering problems~\cite{chen2017exceptional}.

Since finding the coalescence of eigenvectors by numerical full diagonalization can be a tedious work, theoretical development in detecting the EP becomes interesting.
Here the fidelity susceptibility is used to detect these interesting points. Although there are few non-Hermitian generalizations of fidelity suggested in the literature very recenlty~\cite{wrong_def_sum,wrong_def_RR}, we propose a more natural generalization by taking the aforementioned Hilbert space geometry into account.

\section{Generalized fidelity susceptibility}
In this paper,
the fidelity is generalized for comparing both the eigenstates and the metrics in the non-Hermitian Hamiltonian 
with one tunable parameter, e.g. $H(\lambda)=H_0+{\lambda}H'$, with one of the left and right eigenvector pair, $\ket{L(\lambda)}$ and $\ket{R(\lambda)}$, with $\braket{L(\lambda)}{R(\lambda)}=1$. 
We define the generalized (metricized) fidelity 
\begin{equation}\label{eq:fidelity}
        \mathcal{F}:=\Braket{R(\lambda)}{R(\lambda+\epsilon)}
        \Braket{R(\lambda+\epsilon)}{R(\lambda)},
\end{equation}
where $\Bra{R(\lambda)}=\bra{R(\lambda)}G(\lambda)$ and $\Bra{R(\lambda+\epsilon)}=\bra{R(\lambda+\epsilon)}G(\lambda+\epsilon)$.
Although this definition does not necessarily satisfy the first Josza's axiom due to the geometry deformation in the Hilbert spaces,
it is easy to see that when $H=H^\dagger$, $G=\openone$ is Hermitian, positive-definite, and a solution to Eq.~\eqref{eq:eom}. Therefore, $\mathcal{F}$ reduces back to $\mathcal{F}_\text{h}$ in Hermitian systems. While the generalized fidelity $\mathcal{F}$ compares states not only in different geometry but also in the time evolution, $\mathcal{F}$ is constant in time. Using the Schr\"{o}dinger's equation together with Eq.~\eqref{eq:eom}, the time derivative on the generalized fidelity vanishes trivially,
\begin{equation}
        \frac{\partial}{\partial t}\mathcal{F}=0.
\end{equation}
Therefore, this generalized fidelity can indeed be used to quantify the closeness between the states in different geometries of Hilbert space.
Choosing the metric found in Eq.~\eqref{eq:metric} is equivalent
to compute the left and right eigenstates of the Hamiltonian as long as both $\lambda$ and $\lambda+\epsilon$ are not the EP~\cite{atEP},
\begin{equation}\label{eq:BF}
        \mathcal{F}=\braket{L(\lambda)}{R(\lambda+\epsilon)}\braket{L(\lambda+\epsilon)}{R(\lambda)}.
\end{equation}

The generalized fidelity can be expanded as $\mathcal{F}=1-\chi\epsilon^2+\mathcal{O}(\epsilon^3)$ if the parameter $\lambda$ is not an EP and $\epsilon$ is sufficiently small. 
The generalized fidelity susceptibility $\chi$ can be approximated as
\begin{equation}\label{eq:fsus}
        \chi\approx\frac{1-\mathcal{F}}{\epsilon^2}
\end{equation}

The fact that $\mathcal{F}$ reduces back to $\mathcal{F}_\text{h}$ in Hermitian cases means that the generalized fidelity susceptibility $\chi$ can also be used to detect the phase transition, where the fidelity susceptibility density tends to $+\infty$ at the critical point~\cite{[{Note that some higher order quantum phase transitions fail to be detected by the fidelity susceptibility, see }]Tzeng2008b}.
Moreover, since $\mathcal{F}(\lambda)$ is affected by the difference between the metrics $G(\lambda)$ and $G(\lambda+\epsilon)$, the susceptibility $\chi(\lambda)$ is expected to diverge when the Hilbert space is changing its phase. Furthermore, since the sole reason that $\mathcal{F}(\lambda)$ can sit outside of region $[0,1]$ is the metric difference in the Hilbert space, the ``geometric phase change'' is very likely to occur as $\abs{\mathcal{F}(\lambda)}>1$. 
Since $G(\lambda)$ changes smoothly except at the EPs~\cite{Ju2019}, together with the aforementioned properties, the limit of $\Re\chi(\lambda)$ should tend to negative infinity when $\lambda$ approaches an EP ($\tilde{\lambda}$), i.e.,
\begin{equation}\label{eq:criteria}
        \lim_{\lambda\to\tilde{\lambda}}\Re\chi(\lambda)=-\infty.
\end{equation} 

\section{Examples}
\subsection{a toy model}
Here we provide two examples demonstrating the previous statement or the Eq.~\eqref{eq:criteria}.
The simplest single variable adjustable Hamiltonian with an EP is a $\mathcal{PT}$-symmetric Hamiltonian~\cite{Bender2002},
\begin{equation}
        H(r)=\begin{bmatrix}ir&1\\1&-ir\end{bmatrix},
\end{equation}
where $r\in\mathbb{R}$ is the adjustable parameter. The eigenstates coalesce at $r=\pm1$; namely, $r=\pm1$ is an EP of the Hamiltonian. Moreover, the $\mathcal{PT}$-symmetry is preserving as $\abs{r}<1$; however, $\mathcal{PT}$-symmetry is breaking as $\abs{r}>1$.

In the $\mathcal{PT}$ preserving region ($\abs{r}<1$), the time-evolving right eigenstates are
\begin{align}
        \Ket{\psi_1(t)}&=\frac{\exp(-i\sqrt{1-r^2}t)}{\sqrt{2\cos\alpha}}
             \begin{bmatrix}e^{i\alpha/2}\\e^{-i\alpha/2}\end{bmatrix}, \\
        \Ket{\psi_2(t)}&=\frac{i\exp(i\sqrt{1-r^2}t)}{\sqrt{2\cos\alpha}}
             \begin{bmatrix}e^{-i\alpha/2}\\-e^{i\alpha/2}\end{bmatrix},
\end{align}
where $\cos\alpha=\sqrt{1-r^2}$. By using 
Eq.~\eqref{eq:metric} the corresponding metric is
\begin{equation}
G=\frac{1}{\cos\alpha}\begin{bmatrix}1&-i\sin\alpha\\i\sin\alpha&1\end{bmatrix}.
\end{equation}
In the $\mathcal{PT}$ breaking region ($\abs{r}>1$), the time-evolving right eigenstates are
\begin{align}
        \Ket{\psi_1(t)}&=\frac{\exp(t\Lambda)}{\sqrt{2r\Lambda-2\Lambda^2}}
              \begin{bmatrix}1\\-i(r-\Lambda)	\end{bmatrix},\\
        \Ket{\psi_2(t)}&=\frac{\exp(-t\Lambda)}{\sqrt{2r\Lambda-2\Lambda^2}}
              \begin{bmatrix}i(r-\Lambda)\\1\end{bmatrix},
\end{align}
where $\Lambda=\sqrt{r^2-1}$. Then the corresponding metric becomes
\begin{equation}
G=\frac{1}{2\Lambda}\begin{bmatrix}g_{11}&g_{12}\\g_{21}& g_{22}\end{bmatrix},
\end{equation}
where
$g_{11}=(\Lambda+r)e^{-2t\Lambda}-(\Lambda-r)e^{2t\Lambda}$,
$g_{22}=(\Lambda+r)e^{2t\Lambda}-(\Lambda-r)e^{-2t\Lambda}$,
$g_{12}=-i(e^{2t\Lambda}+e^{-2t\Lambda})$,
$g_{21}=i(e^{2t\Lambda}+e^{-2t\Lambda})$.

In both regions, the expansion of the fidelity $\mathcal{F}(r)=\Braket{\psi_1(r)}{\psi_1(r+\epsilon)}\Braket{\psi_1(r+\epsilon)}{\psi_1(r)}$ to the $\epsilon^2$ order is
\begin{equation}
        \mathcal{F}(r)=1+\frac{1}{4(1-r^2)^2}\epsilon^2+\mathcal{O}(\epsilon^3).
\end{equation}
The fidelity is time-independent, and the fidelity susceptibility is
\begin{equation}\label{eq:chi_toy}
        \chi(r)=\frac{-1}{4(1-r^2)^2}.
\end{equation}
Equation~\eqref{eq:chi_toy} shows that $\chi\to-\infty$ as the parameter approaches to the EP $r\to\pm1$ which agrees with Eq.~\eqref{eq:criteria}.

\subsection{Non-Hermitian SSH model}
As a second example, we consider the non-Hermitian Su-Schriffer-Heeger (SSH) model. The SSH model is a one-dimensional (1D) fermionic chain with a bond-alternating hopping term, which originally describes the electrons moving in the polyacetylene~\cite{SSH_1979}. The SSH model provides the simplest 1D free fermion model for studying the topological quantum materials. The investigation on the SSH model is widely extended, from the alternating bonds spin-1/2 XXZ chain~\cite{Tzeng2016} to 3D topological condensed matter systems, e.g. the Weyl semimetals~\cite{Tzeng2020}. Because of its simplicity and non-trivial topological properties, the SSH model has been treated as a parent model for studying non-Hermitian many-body systems~\cite{wrong_def_sum,Song_2019,nSSH_optical_2018,PhysRevLett.121.086803,pan2018photonic,nSSH_es1,nSSH,ghost}.
Since the experimental realization of the non-Hermitian SSH model is easier to set up in a system with finite number of unit cells with gain and loss, in the following, we mainly focus on the finite system with $N$ unit cells, and linking to the thermodynamic limits. 
The Hamiltonian is
\begin{align}\label{eq:nSSH}
H_\text{SSH}& =v\sum_{n=1}^N(c_{n\uparrow}^\dagger c_{n\downarrow} + \mathrm{H.c.})
            +w\sum_{n=1}^N(c_{n\downarrow}^\dagger c_{n+1\uparrow} + \mathrm{H.c.})\nonumber\\
            & +iu\sum_{n=1}^N(c_{n\uparrow}^\dagger c_{n\uparrow}
              -c_{n\downarrow}^\dagger c_{n\downarrow})
\end{align}
Where the parameters $u$, $v$, $w\in\mathbb{R}$, and $u\geq0$. $c_{n\uparrow}$ and $c_{n\downarrow}$ are the annihilation operators at the $n$th unit cell for the site with gain and loss, respectively. The usual anti-commutation relation, $\{c_{n\sigma},c_{m\tau}^\dagger\}=\delta_{nm}\delta_{\sigma\tau}$, and the periodic boundary conditions, $c_{N+1\sigma}=c_{1\sigma}$, are applied. 
The phase diagram of the non-Hermitian SSH model in the thermodynamic limits has been studied. The system is in the $\mathcal{PT}$ symmetric topological phase $w>v+u$, the $\mathcal{PT}$ broken phase $v-u<w<v+u$, and the $\mathcal{PT}$ symmetric trivial phase $w<v-u$, respectively~\cite{nSSH,ghost}.
By employing Fourier transform, $\tilde{c}_{k\sigma}=\frac{1}{\sqrt{N}}\sum_{n=1}^Ne^{ikn}c_{n\sigma}$, where $k=2\pi m/N$ and $m=0,\dots,N-1$. The Hamiltonian becomes
\begin{equation}
        H_\text{SSH}=\sum_{k}
        \begin{bmatrix}\tilde{c}_{k\uparrow}^\dagger & \tilde{c}_{k\downarrow}^\dagger\end{bmatrix}
        \begin{bmatrix} iu & \xi_k \\ \xi_k^* & -iu \end{bmatrix}
        \begin{bmatrix}\tilde{c}_{k\uparrow} \\ \tilde{c}_{k\downarrow} \end{bmatrix},
\end{equation}
where $\xi_k=v+we^{ik}$. The eigenvalues for the single particle is $\varepsilon^{\pm}_k=\pm\sqrt{\abs{\xi_k}^2-u^2}$.
After constructing the set of operators
$(\psi^{\text{R}\pm}_k)^\dagger=(\varepsilon^\pm_k+iu)\frac{\xi_k}{\abs{\xi_k}^2}\tilde{c}_{k\uparrow}^\dagger+\tilde{c}_{k\downarrow}^\dagger$, and
$\psi^{\text{L}\pm}_k=\frac{1}{2\varepsilon^\pm_k}[\xi_k^*\tilde{c}_{k\uparrow}+(\varepsilon^\pm_k-iu)\tilde{c}_{k\downarrow}]$,
the Hamiltonian becomes diagonalized to
\begin{equation}\label{eq:diagH}
        H_\text{SSH}=\sum_k\left[
                     \varepsilon^+_k(\psi^{\text{R}+}_k)^\dagger\psi^{\text{L}+}_k
                    +\varepsilon^-_k(\psi^{\text{R}-}_k)^\dagger\psi^{\text{L}-}_k\right].
\end{equation}
Where $\psi^{\text{R}\pm}_k$ and $\psi^{\text{L}\pm}_{k'}$ satisfy 
$\bra{0}(\psi^{\text{R}\pm}_k)^\dagger=0=\psi^{\text{L}\pm}_k\ket{0}$, $\ket{0}$ is the vacuum state, and $\braket{0}{0}=1$. Moreover, their (anti-)commutators with the Hamiltonian are
$[H_\text{SSH},(\psi^{\text{R}\pm}_k)^\dagger]=\varepsilon^{\pm}_k(\psi^{\text{R}\pm}_k)^\dagger$,
$[H_\text{SSH},\psi^{\text{L}\pm}_k]=\varepsilon^{\pm}_k\psi^{\text{L}\pm}_k$,
and $\lbrace\psi^{\text{L}\tau}_k,(\psi^{\text{R}\tau'}_{k'})^\dagger\rbrace=\delta_{kk'}\delta^{\tau\tau'}$.


The time-dependent metric tensor can be chosen to be
\begin{widetext}
\begin{align}
G(t)=\sum_{p,q=0}^N\sum_{k,k'}\left(\prod_{a=1}^pe^{-2t\Im\varepsilon^+_{k_a}} \right)\left(\prod_{b=1}^qe^{-2t\Im\varepsilon^-_{k'_b}}\right)\left(\prod_{a=1}^p\psi^{\text{L}+}_{k_a}\right)^\dagger\left(\prod_{b=1}^q\psi^{\text{L}-}_{k'_b}\right)^\dagger\ket{0}\bra{0}\left(\prod_{b=1}^q \psi^{\text{L}-}_{k'_b}\right)\left(\prod_{a=1}^p\psi^{\text{L}+}_{k_a}\right),
\end{align}
\end{widetext}
where
\begin{align*}
\sum_{k,k'}\equiv  \sum_{k_1=1}^N\sum_{k_2=k_1+1}^N...\sum_{k_p=k_{p-1}+1}^N 
 \sum_{k'_1=1}^N\sum_{k'_2=k'_1+1}^N...\sum_{k'_q=k'_{q-1}+1}^N.
\end{align*}






\begin{figure}
\centering
\includegraphics[width=3.3in]{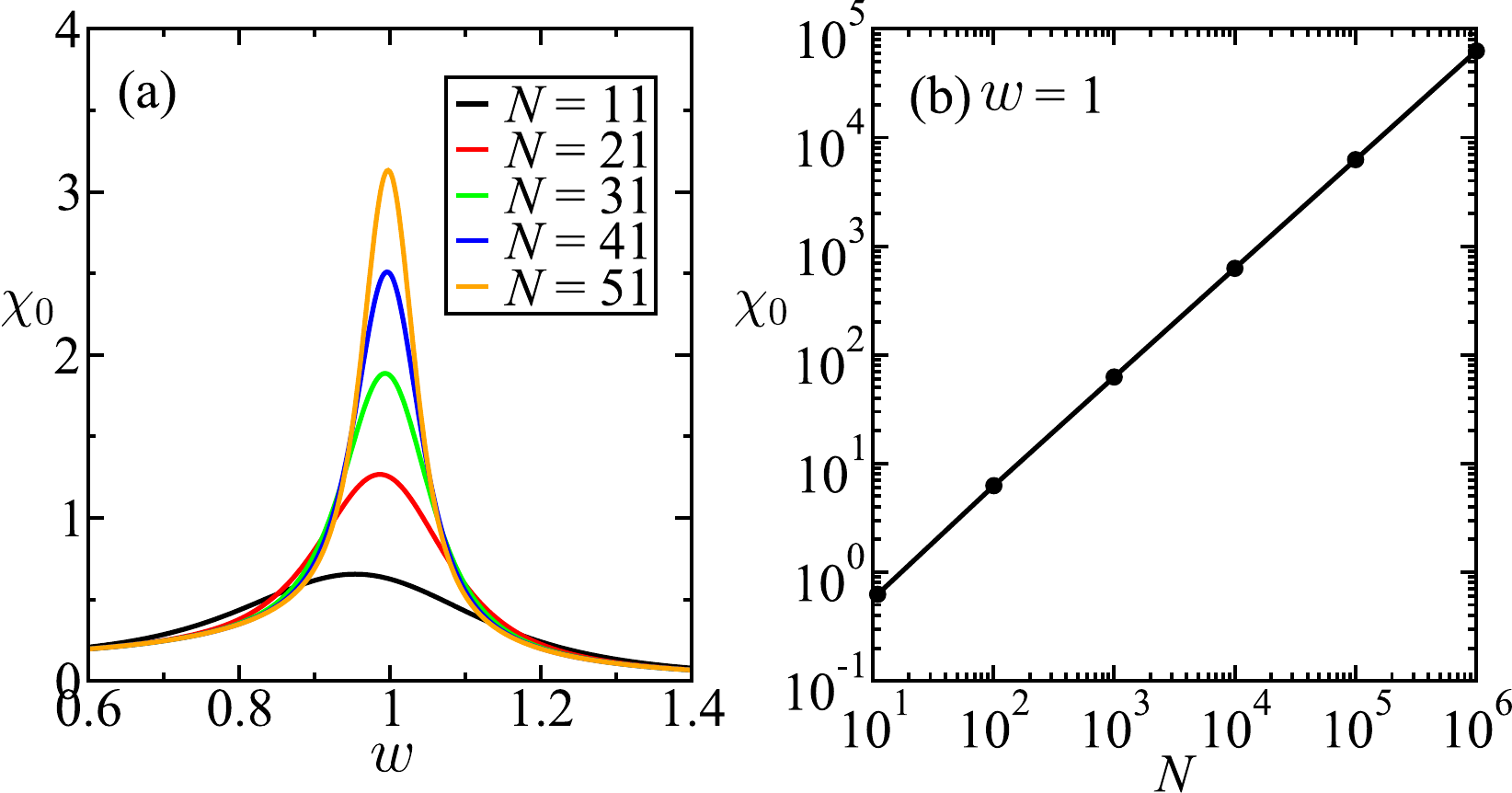}
\caption{Ground state fidelity susceptibility density $\chi_0$ for the Hermitian ($u=0$) SSH model. $v=1$ is set as the unit. (a) $\chi_0$ as a function of $w$. (b) The size dependence at the quantum critical point $w=1$ is plotted, and a power-law divergence is found, $\chi_0=5(N-1)/80$.}\label{fig:fsus_u0}
\end{figure}
From the Eq.~\eqref{eq:diagH}, we define that the half-filled right eigenstate $\Ket{\Psi_0}$ such that $(\psi^{\text{R}-}_k)^\dagger\Ket{\Psi_0}=0$ for all $k$ is the ``ground state''. In other words, the ground state is the state full of particles in the $\varepsilon_k^-$ band, even $\varepsilon_k^-$ can be complex in the $\mathcal{PT}$ broken phase.
The time dependent right ground state is
\begin{equation}
        \Ket{\Psi_0(t)}=\left[\prod_k e^{-it\varepsilon_k^-}\left(\psi^{\text{R}-}_k\right)^\dagger \right]\ket{0},
\end{equation}
and the corresponding dual vector of the right ground state, $\Bra{\Psi_0(t)}=\bra{\Psi_0(t)}G(t)$ is the left ground state,
\begin{equation}
        \Bra{\Psi_0(t)}=\bra{0}\left(\prod_k e^{it\varepsilon_k^-}\psi^{\text{L}-}_k \right).
\end{equation}
The fidelity susceptibility density $\chi_0$ for the two ground states with parameters $(u,v,w)$ and $(u,v,w+\epsilon)$ is
\begin{equation}\label{eq:fsus_SSH}
        \chi_0=\frac{1}{N}\sum_k\frac{v^2\sin^2k-u^2}{4(v^2+w^2+2vw\cos k-u^2)^2}, 
\end{equation}
where $\epsilon$ is taken to be infinitesimally small. It is clear that $\chi_0$ is a time independent quantity.

In the case $u=0$, the Hermitian limit, the ground state fidelity susceptibility density Eq.~\eqref{eq:fsus_SSH} for $v=1$ is plotted in Fig.~\ref{fig:fsus_u0}. For $N$ is even, there is a $k=\pi$ such that Eq.~\eqref{eq:fsus_SSH} diverges at the critical point $(u,v,w)=(0,1,1)$. Therefore we focus on the odd $N$ for looking at the size dependences, where $\chi_0$ is always finite. It is known that $\chi_0$ goes to $+\infty$ at the quantum critical point $w=v$ in the thermodynamic limit $N\to\infty$. The scaling properties of the fidelity susceptibility has been studied for the quantum phase transitions~\cite{Tzeng2008a}. From Eq.~\eqref{eq:fsus_SSH} we obtain a power-law divergence $\chi_0=5(N-1)/80$ at $v=w=1$, as shown in Fig.~\ref{fig:fsus_u0}(b).

\begin{figure}
\centering
\includegraphics[width=3.3in]{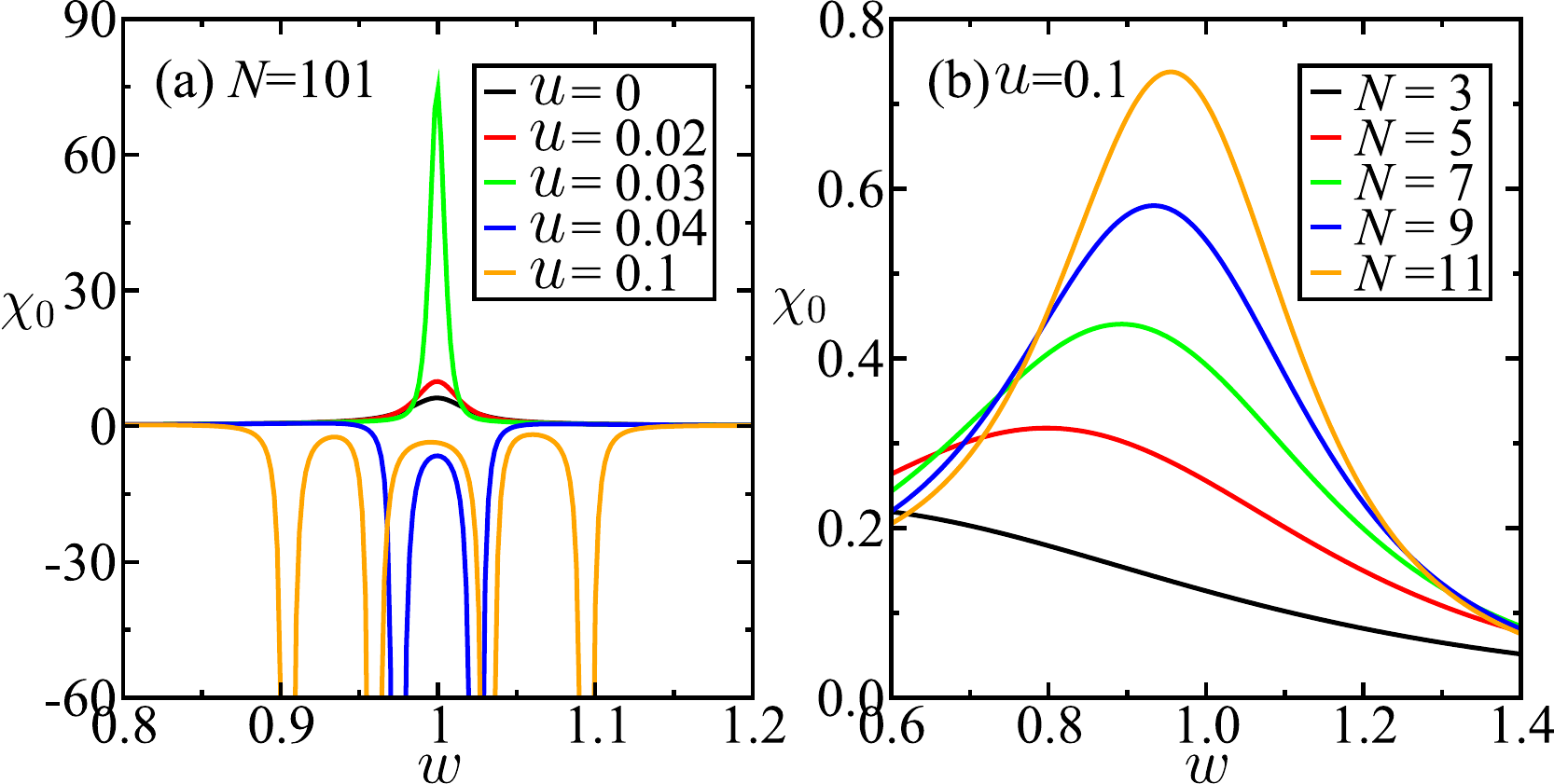}
\caption{Ground state fidelity susceptibility density $\chi_0$ for the non-Hermitian SSH model. $v=1$ is set as the unit. (a) For $N=101$, the signal of the Hermitian quantum phase transition is enhanced by adding a small non-Hermitian term. The EP is not present in the finite $N=101$ until $u\gtrsim0.04$.
(b) For $u=0.1$, small finite size $N$ would lead the absence of the EP.}\label{fig:fsus_N101}
\end{figure}

For the non-Hermitian ($u\neq0$) case, as $\abs{w-v}\leq\abs{u}$, there is a $k=k_\text{EP}$ such that the denominator vanishes. 
Since the denominator is always positive and the numerator is always negative, the ground state fidelity susceptibility density $\chi_0\to-\infty$ characterizes an EP if one of the $k$ points closes to the $k_\text{EP}$. Therefore in the thermodynamics limits, i.e. the values of $k$ are continuous in $[0,2\pi)$, the $\mathcal{PT}$ broken phase $v-u\leq w\leq v+u$ becomes the exceptional line. In other words, the exceptional line is an emergent phenomena in the thermodynamics limits, while in the finite system the exceptional line breaks down into separate EPs. For example, as shown in Fig.~\ref{fig:fsus_N101}(a), the exceptional line in the finite size system $N=101$ breaks down into 2 EPs and 4 EPs for $u=0.04$ and $u=0.1$, respectively. As the parameter approaching to these EPs, the behavior of $\chi_0$ agrees with the hypothesis Eq.~\eqref{eq:criteria}.


Interestingly,
since there is only a limited number of $k$ points in the finite size system, we find no EP at all in the whole parameter space if the non-Hermitian driving parameter $u$ is small enough or the size $N$ is odd and small.
As shown in Fig.~\ref{fig:fsus_N101}(a) for larger size $N=101$ and smaller sizes in Fig.~\ref{fig:fsus_N101}(b), the ground state fidelity susceptibility density $\chi_0$ is always positive and finite for $u\lesssim0.03$ with $N=101$ system and $u=0.1$ for smaller systems, showing no EP in the whole parameter space.
The absence of EP means that these finite size systems are $\mathcal{PT}$ preserving in the whole $(v,w)$ parameter region, i.e. the $\mathcal{PT}$ breaking phase vanishes in the odd $N$ systems.
Although the terminology ``phases of matter'' implies the case is in the thermodynamic limits, the system with size $N=101$ is usually large enough for distinguishing different phases in many numerical studies about phase transitions. 
Our result shows that $\mathcal{PT}$ breaking region does not exist for finite size systems provide $u$ is small enough, which further extends the discussions in the previous findings~\cite{nSSH,ghost}.
For example $u=0.1$, the small size systems in Fig.~\ref{fig:fsus_N101}(b) have no EP. However once the size is large enough, at least two EPs present at the phase boundary, as shown in Fig.~\ref{fig:fsus_N101}(a). For $u\leq0.03$, the size should be much larger than $N=101$ for opening the $\mathcal{PT}$ breaking region.
Because the sensitivity of the non-Hermitian system could be enhanced near the EP~\cite{Nature2017,chen2017exceptional},
the small $u$ which is not strong enough to open the $\mathcal{PT}$ breaking region, consequently, enhances the signal of the phase transitions between the topological phase and the trivial phase. As shown in Fig.~\ref{fig:fsus_N101}(a), $\chi_0$ with small $u$ becomes much larger than the case $u=0$. 
The amplified susceptibility means that by a small non-Hermitian perturbation, the finite size system becomes more sensitive in responding the quantum phase transition. 
In other words, we present an example that a small non-Hermitian perturbation is helpful to detect quantum phase transitions. For those with particular difficulty, e.g. higher-order quantum phase transitions~\cite{Tzeng2008b} and the Kosterlitz-Thouless transition~\cite{Kosterlitz_1973, MFYang2007, fsus_BKT}, further investigations on the generalized fidelity susceptibility are interesting and remaining open questions.

\section{conclusion}
We have properly generalized the fidelity Eq.~\eqref{eq:fidelity} for non-Hermitian quantum systems by including the metric of the Hilbert space. We mention that, unlike the phase transition, the EP is not a thermodynamic but the geometric property of the Hilbert space. 
In the perspective of the fidelity susceptibility, both the EP and the quantum critical point enhance the sensitivity of a non-Hermitian system, however the EP is a geometric effect and divergence direction is clearly distinguished from the quantum critical point.
We propose Eq.~\eqref{eq:criteria} is one of the basic properties of EPs.
Since the previous study of the fidelity shows the failure of detecting some quantum phase transitions~\cite{Tzeng2008b}, this hypothesis for the EPs might also break down in some non-Hermitian systems.
However, we have not found any counterexample against Eq.~\eqref{eq:criteria}.

Moreover, since the computation of the generalized fidelity is reduced to the biorthogonal formalism Eq.~\eqref{eq:BF} if the parameter is not the EP~\cite{atEP}, numerical computations for more complicated systems require only one of the biorthogonal eigenpair of the non-Hermitian Hamiltonian, instead of solving the metric equation of motion Eq.~\eqref{eq:eom} or full diagonalization.

\begin{acknowledgments}
YCT is grateful to Min-Fong Yang and Po-Yao Chang for many useful discussions.
YCT acknowledges support by the Ministry of Science and Technology of Taiwan (MOST) under grants No. MOST 108-2811-M-005-522, 108-2112-M-029-002, and 109-2636-M-007-003.
WMH and CYJ are supported by the MOST through 107-2112-M-005-008-MY3 and 109-2811-M-005-509, respectively.
GYC and CYJ are supported partially by the MOST, through grant No. MOST 109-2112-M-005-002.
\end{acknowledgments}

%


\begin{thebibliography}{57}
\bibitem [{atE()}]{atEP}%
  \BibitemOpen
  \href@noop {} {}\bibinfo {note} {The definition of the generalized fidelity
  Eq.~\eqref{eq:fidelity} still holds at the EP, however one should solve the
  equation of motion Eq.~\eqref{eq:eom} for finding a valid metric. The
  biorthogonal fidelity in Eq.~\eqref{eq:BF} is only convenient for calculation
  outside the EP. At the EP, it loss the geometric meaning of the Hilbert
  space. Here we investigate the behavior of the biorthogonal fidelity
  suseptibility $\chi$ as the parameter approaches to the EP.}\BibitemShut
  {Stop}%
\end{thebibliography}
\end{document}